\documentclass[amsmath,amssymb,a4paper,twocolumn,prl]{revtex4}
\usepackage{amssymb}
\usepackage{amsmath}
\usepackage{bm}
\usepackage{txfonts}
\usepackage{graphicx}
\usepackage{ulem}
\usepackage{color}
\usepackage[colorlinks=true, allcolors=blue]{hyperref}

\usepackage{bm}
\usepackage{array}

\begin{document}

\title{One-way topological edge states in nonlinear gyroscopic phononic crystals}
\author{Jingtian \surname{Shi}}

\affiliation{
Department of Physics, Tsinghua University, Beijing 100084, People's Republic of China
}

\begin{abstract}

A unified form of time-reversal symmetry (TRS) breaking terms in phononic crystals, leading to nontrivial phononic topology, has been proposed recently, but is contradicted by some other works which introduce gyroscopic effect as TRS-breaking. We re-study gyroscopic phononic crystals using Newtonian mechanical method, and find the correct TRS-breaking term in consistent with the unified form. Applying this term we calculate the basic topological phononics in a honeycomb lattice. Furthermore, we study nonlinear phonon-phonon scattering effect on topological phononic edge states by molecular dynamics simulation. Generally edge states are not immune to such scattering effect, but under specific conditions some edge states run into bulk much more slowly, depending on the parameters of the model. This opens up the potential for effectively suppressing phonon dissipation by tuning the parameters, thereby realizing near-100\% efficiency of one-way phonon transport in phonon devices.

\end{abstract}
\maketitle

\section{I. Introduction}
The discovery of topological quantum states of matter, like the quantum (anomalous/spin) Hall [Q(A/S)H] effect, represents one of the most important achievements in condensed matter physics \cite{Hasan2010, Qi2011}. The concept of topology was originally developed for electrons and has recently been generalized for phonons, leading to findings of novel topological phononic states and unprecedentedly new ways of controlling phonons~\cite{Liu2017,Kariyado2015}. For instance,  the QAH-like states characterized by nonzero topological invariant Chern number were proposed in phononic crystals with broken time reversal symmetry (TRS)~\cite{Kariyado2015,Lifa2010,YT2015}. This kind of exotic phononic states give one-way phononic edge states immune to elastic scattering, which could find important applications in low-dissipation phononic conduction and high-efficient phononic diode~\cite{YL2017}.

Various mechanisms have been proposed to break the TRS of phononic crystals, including Coriolis field, magnetic field in ionic lattices, and spin-lattice coupling in magnetic
materials~\cite{Lifa2010}. A recent work demonstrated that the TRS-breaking interactions, when the nonlinear (or anharmonic) part is neglected, can be described in a unified form by a harmonic Lagrangian $L' =  \eta_{ij} \dot{u}_i u_j$, where $\dot{u}_i$ and $u_j$ are the velocity and displacement of the $i^{\text{th}}$ and $j^{\text{th}}$ degree of freedom, respectively. Importantly, the coefficient matrix $\eta$ is real and antisymmetric, so that the Hamiltonian can be Hermitian for ensuring real eigenvalues and probability conservation~\cite{YL2017}. The argument is in agreement with almost all previous results~\cite{Kariyado2015,Lifa2010,YT2015}. As the only known exception, Refs.~\cite{P2015,CARTA2014,Brunrspa2012} introduced gyroscopic coupling to break TRS, but their TRS-breaking effect was included by an imaginary mass term or equivalently described by another type of Lagrangian $L'' =  \alpha_{ij} \dot{u}_i \dot{u}_j$, where the coefficient matrix $\alpha$ is imaginary and antisymmetric, in contradict with the previous argument. This raises a problem:  Either the argument of $L' $ is not general enough to take gyroscopic coupling into consideration or the result of $L''$ is questionable. In either case, it is worth to revisit the TRS-breaking effect of gyroscopic phononic crystals that were shown to have the QAH-like states by Refs.~\cite{P2015,CARTA2014,Brunrspa2012}. Furthermore, gyroscopic coupling was treated in the harmonic approximation in the previous work. The nonlinear interactions that were usually neglected in most studies are well known to be important in determining transport properties of bulk phonons. Their influence on the one-way phononic edge states, however, remains largely unknown.

In this work, we re-derive the dynamic equations and the TRS-breaking effect of gyroscopic phononic crystals. We find that the correct TRS-breaking term is Coriolis-like, thus allowed by the argument of $L'$. Applying this term into a honeycomb lattice, we calculate different topological orders and one-way phononic edge states induced by various TRS-breaking field strengths, as previously done by Y. T. Wang et al~\cite{YT2015} in Coriolis field, and draw similar conclusions. To investigate nonlinear effect on one-way phononic edge states, we carry out molecular dynamics simulations based on exact nonlinear dynamic equations of our gyroscopic honeycomb lattice model, in which we observe the scattering process of edge states. We see different patterns of scattering process for different edge states: within certain time scales, some can be scattered into the bulk, some are scattered mainly into other edge states, and others are relatively robust against scattering. By tuning the parameters we can selectively make certain edge states robust, thus optimal for applications in phonon devices.

\section{II. Model of gyroscopic phononic crystals}

\begin{figure}[t]
    \centering
    \includegraphics[width=0.48\textwidth,angle=0]{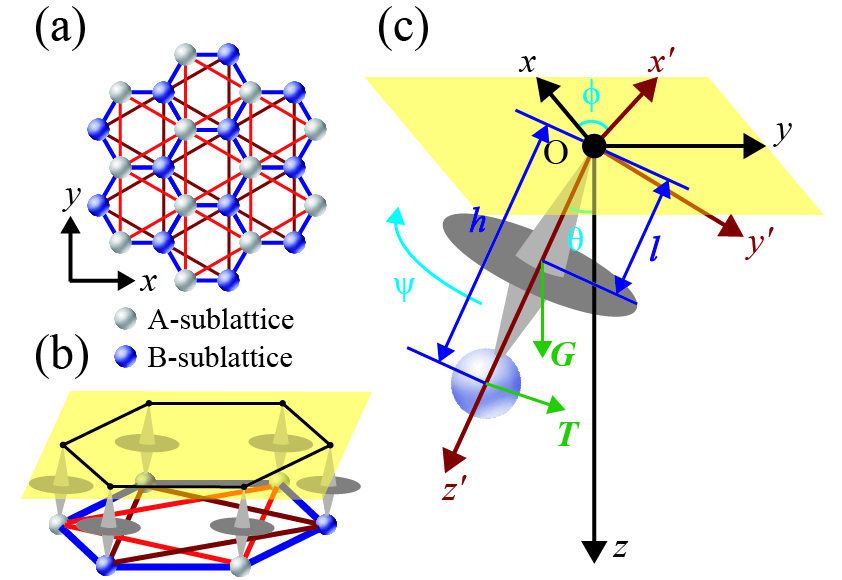}
    \caption{(a) A gyroscopic phononic crystal in a honeycomb lattice, in which the nearest-neighbor and next-nearest-neighbor interactions are considered and indicated by blue and (dark) red lines, respectively. (b) Each site is composed of a gyroscope that rotates around a fixed point. The fixed points (denoted by the black dots) form a honeycomb structure in the horizontal $xy$-plane (shaded yellow). (c) Structural details of a gyroscope. $\theta$, $\phi$, and $\psi$ are the Euler angles. $h$ and $l$ indicate the distances to the fixed point $O$. $\bm{T}$ and $\bm{G}$ denote respectively the total spring force and gravity acting on the gyroscope. The coordinate frame $O-x'y'z'$ is obtained by rotating $O-xyz$ about $z$-axis by $\phi$ and then about $x'$-axis by $\theta$, guaranteeing that $x'$-axis always lies in $xy$-plane and $z'$-axis always coincides with the axis of gyroscope.}
    \label{structure}
\end{figure}

For a general 2D phononic crystal we introduce gyroscopic coupling by attaching identical gyroscopes to each site. Each gyroscope can rotate around a fixed point and all fixed points form a 2D lattice corresponding to the specific crystal, lying in $xy$-plane. Site-site interactions are represented by spring forces between tips of gyroscopes. Figure \ref{structure}(a) and (b) show an example of gyroscopic honeycomb lattice with the nearest-neighbor (NN) and next-nearest-neighbor (NNN) interactions considered. The orientation of each gyroscope is specified by Euler angles $(\phi,\theta,\psi)$, as shown in Fig. \ref{structure}(c). This model differs from those studied by Refs.~\cite{P2015,CARTA2014,Brunrspa2012} only in one aspect: we consider the gravity of gyroscope and choose $z$-axis to be downwards to make the equilibrium positions stable.

For arbitrary motion of a gyroscope, the angular velocity can be divided into three components: $\bm{\omega}=\bm{\omega}_p+\bm{\omega}_n+\bm{\omega}_s$, where $\bm{\omega}_p=\dot{\phi}\hat{z}$, $\bm{\omega}_n=\dot{\theta}\hat{x}'$ and $\bm{\omega}_s=\dot{\psi}\hat{z}'$ represent precession, nutation and spin, respectively. The angular momentum with respect to the fixed point $O$ is given by $\bm{L}=\bm{I\omega}$, where $\bm{I}$ is the moment of inertia tensor of the gyroscope. Gyroscopes considered in this study are rotationally symmetric. In the $x'y'z'$ coordinate, $\bm{I}=\text{diag}\{I_1,I_2,I_3\}$ ($I_1=I_2=I_0$) and the total moment acting on the gyroscope with respect to $O$ is $\bm{N}={\rm d}\bm{L}/{\rm d}t$, which is expressed as:
\begin{equation}
    \left\{\begin{array}{>{\displaystyle}l}
        N_{x'}=I_0(\ddot{\theta}-\dot{\phi}^2\sin{\theta}\cos{\theta})+I_3\dot{\phi}\sin\theta(\dot{\phi}\cos{\theta+\dot{\psi}})\\
        N_{y'}=I_0(\ddot{\phi}\sin\theta+2\dot{\phi}\dot{\theta}\cos\theta)-I_3\dot{\theta}(\dot{\phi}\cos{\theta+\dot{\psi}})\\
        N_{z'}=I_3(\ddot{\phi}\cos\theta-\dot{\phi}\dot{\theta}\sin\theta+\ddot{\psi})
    \end{array}\right.
    \label{myN}
\end{equation}

In comparison, Ref.~\cite{Brunrspa2012} gave a different result for the same problem:
\begin{equation}
    \left\{\begin{array}{>{\displaystyle}l}
        N_{x'}=I_0(\ddot{\theta}-\dot{\phi}^2\sin{\theta}\cos{\theta})+I_3\dot{\phi}\sin\theta(\dot{\phi}\cos{\theta+\dot{\psi}})\\
        N_{y'}=\sin\theta[I_0(\ddot{\phi}\sin\theta+2\dot{\phi}\dot{\theta}\cos\theta)-I_3\dot{\theta}(\dot{\phi}\cos{\theta+\dot{\psi}})]\\
        N_{z'}=I_3(\ddot{\phi}\cos\theta-\dot{\phi}\dot{\theta}\sin\theta)
    \end{array}\right.
    \label{theirN}
\end{equation}
There are two major contradictions compared to our result: (i) An extra factor $\sin\theta$ is added in $N_{y'}$, which we believe is wrong. The reason is that for small $\theta$, $N_{x'}$ is a first-order small quantity, while $N_{y'}$ would be a second-order small quantity if the extra factor $\sin\theta$ were included. This contradicts with the symmetry requirement that $N_{x'}$ and $N_{y'}$ should be of the same order. (ii) $\ddot{\psi}$ is neglected in $N_{z'}$ by assuming a constant spinning rate $\dot{\psi}$. However, the total moment of the spring force $\bm{T}$ and gravity force $\bm{G}$ is given by $\bm{N}=\bm{h}\times\bm{T}+\bm{l}\times\bm{G}$. $N_{z'}=0$ is ensured by the fact that $\bm{h} = h \hat{z}'$ and $\bm{l} = l \hat{z}'$. Thus, it is $\Psi=\dot{\phi}\cos\theta+\dot{\psi}$ instead of $\dot{\psi}$ that is a time independent constant. After eliminating the variable $\psi$ by the constant condition,
Eq. (\ref{myN}) becomes:
\begin{equation}
    \left\{\begin{array}{>{\displaystyle}l}
        N_{x'}=I_0(\ddot{\theta}-\dot{\phi}^2\sin{\theta}\cos{\theta})+I_3\Psi\dot{\phi}\sin\theta\\        N_{y'}=I_0(\ddot{\phi}\sin\theta+2\dot{\phi}\dot{\theta}\cos\theta)-I_3\Psi\dot{\theta}
    \end{array}\right.
    \label{myNp}
\end{equation}
In contrast, Refs.~\cite{Brunrspa2012,P2015} employed a constant condition $N_x=N_y=0$, despite the fact that $N_x$ and $N_y$ are obviously nonzero and crucial to define the equation of motion.

Now we consider the motion of the tip in $xy$-plane. Let $\bm{U}=(U_x,U_y)^T$ be the in-plane displacement i.e. the in-plane projection of $\bm{h}$. Following Eq. (\ref{myNp}) we derive the in-plane dynamic equations without approximation in the supplementary material (SM)~\cite{SM}. The exact dynamic equations are:
\begin{equation}
    \left\{\begin{array}{>{\displaystyle}l}
        I_0\frac{\rm d}{{\rm d}t}\left(z^2\frac{\rm d}{{\rm d}t}\frac{x}{z}\right)+I_3\Psi\dot{y}+h\left(xT_z-zT_x\right)+lGx=0\\
        I_0\frac{\rm d}{{\rm d}t}\left(z^2\frac{\rm d}{{\rm d}t}\frac{y}{z}\right)-I_3\Psi\dot{x}+h\left(yT_z-zT_y\right)+lGy=0
    \end{array}\right.
    \label{strictdyn}
\end{equation}
where $x=U_x/h$, $y=U_y/h$, $z=U_z/h=\sqrt{1-x^2-y^2}$ and $T_x$, $T_y$ and $T_z$ are components of $\bm{T}$. Note that under small-$\theta$ approximation $x$, $y$ and $T_z$ are small quantities and $z\approx1$, the approximate dynamic equation is given by
\begin{equation}
	\ddot{\bm{U}}+2\left(\begin{array}{cc}
    0&-\Omega\\\Omega&0
    \end{array}\right)\dot{\bm{U}}+g\bm{U}-\frac{\bm{T}}{M}=0
    \label{appdyn}
\end{equation}
where $\bm{T}$ is the in-plane projection of the total spring force acting on the site, $\displaystyle M=\frac{I_0}{h^2}$, $\displaystyle g=\frac{lG}{I_0}$ and $\displaystyle\Omega=-\frac{I_3\Psi}{2I_0}$ are constants.

Note that the spring force is a linear response to the displacement under small-$\theta$ approximation, i.e. $\bm{T}=-\mathbf{C}\bm{U}=-M\mathbf{D}\bm{U}$ where $\mathbf{C}$ and $\mathbf{D}$ are respectively the force constant matrix and the dynamic matrix, a Fourier transform of Eq. (\ref{appdyn}) in a unit cell containing two sublattices yields the phononic secular equation:
\begin{equation}
	\left(-\omega_{\bm{k}}^2-2{\rm i}\omega_{\bm{k}}\pmb{\eta}+\tilde{\mathbf{D}}(\bm{k})\right)\bm{u_k}=0
    \label{secular}
\end{equation}
where $\pmb{\eta}=\left(\begin{array}{cccc}
	0&-\Omega&0&0\\
    \Omega&0&0&0\\
    0&0&0&-\Omega\\
    0&0&\Omega&0
\end{array}\right)$ is the TRS-breaking term induced by gyroscopic effect, $\tilde{\mathbf{D}}(\bm{k})=\mathbf{D}(\bm{k})+g$ is the $4\times4$ net dynamic matrix in $\bm{k}$-space, and $\bm{u_k}=\left(u_{Ax},u_{Ay},u_{Bx},u_{By}\right)$ is the in-plane displacement of $A$ and $B$ sites within a unit cell. We notice immediately that this is basically the same as the secular equation in Coriolis field proposed before~\cite{YL2017,YT2015} except that the finite weight of gyroscope is taken into account here. Obviously our secular equation is compatible with the general discussion of TRS-broken harmonic dynamics given by Y. Liu et al~\cite{YL2017}, according to whom the TRS-breaking phononic secular equation of a harmonic lattice must take the form $\left(-\omega_{\bm{k}}^2-2{\rm i} \omega_{\bm{k}} \pmb{\eta}(\bm{k}) +\mathbf{D}(\bm{k}) \right)\bm{u_k}=0$
where the TRS-breaking matrix $\pmb{\eta}(\bm{k})$ is \textit{anti}-Hermitian and the dynamical matrix $\mathbf{D}(\bm{k})$ is Hermitian and semi-positive definite.

All of the gyroscopic models studied in References~\cite{P2015,CARTA2014,Brunrspa2012} fit our general model in the case that $g=0$. However, their version of secular equation is
\begin{equation}
    \left(-\omega_{\bm{k}}^2\tilde{\mathbf{M}}+\mathbf{C}(\bm{k})\right)\bm{u_k}=0
    \label{their_dyn}
\end{equation}
where $\tilde{\mathbf{M}}=\left(\begin{array}{cccc}
    M&-{\rm i}\alpha&0&0\\
    {\rm i}\alpha&M&0&0\\
    0&0&M&-{\rm i}\alpha\\
    0&0&{\rm i}\alpha&M
\end{array}\right)$ is their ``TRS-broken mass matrix'' and $\mathbf{C}(\bm{k})$ is the $\bm{k}$-space force constant matrix. The essential difference is that their TRS-breaking term is proportional to $\omega_{\bm{k}}^2$ rather than $\omega_{\bm{k}}$, which is excluded by the discussion in Ref.~\cite{YL2017}. We believe that Eq. (\ref{their_dyn}) is physically not allowed because it breaks the ``particle-hole'' symmetry: if $(\bm{k},\omega_{\bm{k}},\bm{u_k})$ is a solution, then $(-\bm{k},-\omega_{\bm{k}},\bm{u_k}^*)$ is not a solution, which prevents real-space displacements from keeping real. On the other hand, one can easily check that Eq. (\ref{secular}) preserves the ``particle-hole'' symmetry.

\section{III. Topological orders of honeycomb lattice}

Now that we have fitted our gyroscopic model for Ref.~\cite{YL2017}, we can employ their method to study the phononic band topology: to transform Eq. (\ref{secular}) equivalently into a Schr\"{o}dinger-like equation $H(\bm{k})\psi_{\bm{k}}=\omega_{\bm{k}}\psi_{\bm{k}}$, where
\begin{equation}
	H(\bm{k})=\left(\begin{array}{cc}
    	0 & {\rm i}\mathbf{D}^{1/2}(\bm{k})\\
        -{\rm i}\mathbf{D}^{1/2}(\bm{k}) & -2{\rm i}\pmb{\eta}
    \end{array}\right),\quad
    \psi_{\bm{k}}=\left(\begin{array}{c}
    	\mathbf{D}^{1/2}(\bm{k})\bm{u_k}\\
        -{\rm i}\omega_{\bm{k}}\bm{u_k}
    \end{array}\right)
\end{equation}
and to define \textit{Berry curvature} and \textit{Chern number} in analogy to quantum mechanics~\cite{SM}. For each band gap, the \textit{gap Chern number} is defined as the sum of Chern numbers over all bands below the gap, to verify the topolgical nature of the gap.

\begin{figure*}[t]
    \centering
    \includegraphics[width=0.96\textwidth,angle=0]{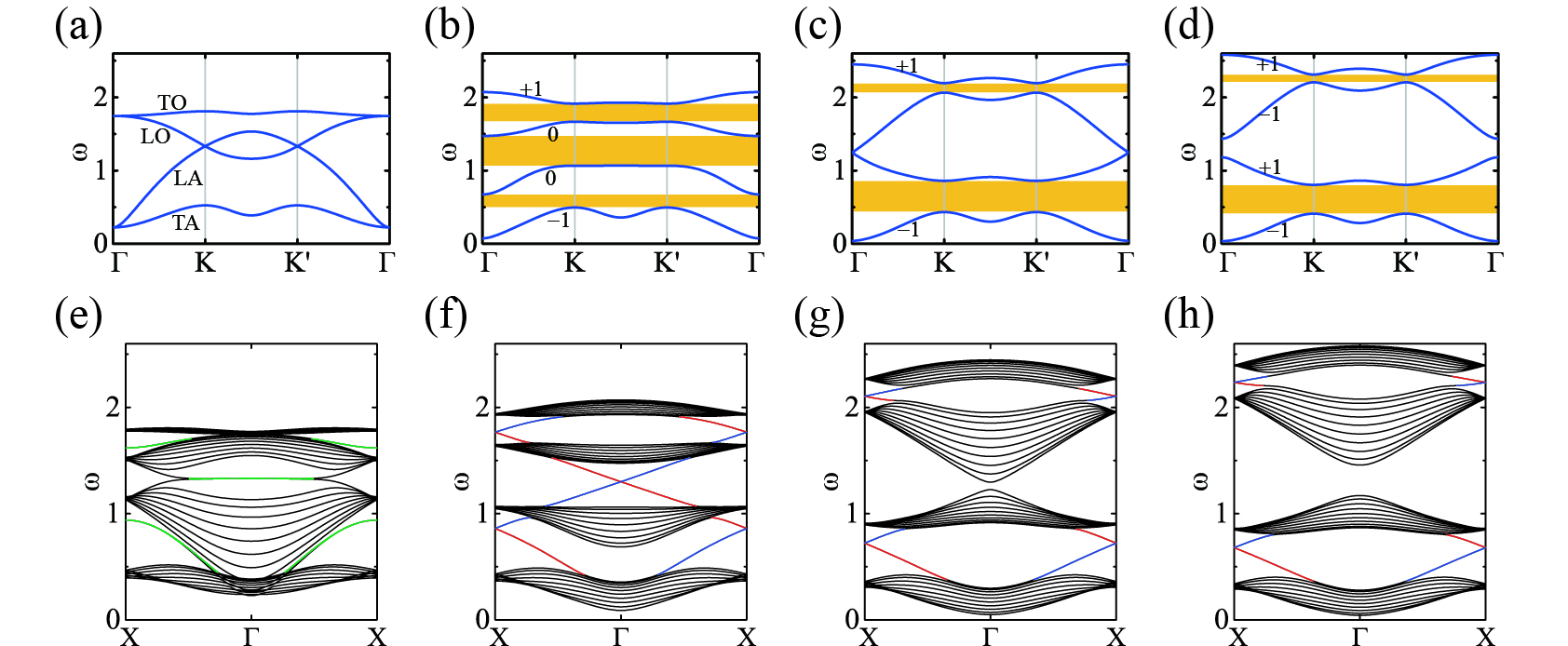}
    \caption{(a)-(d) The phononic dispersion of gyroscopic honeycomb lattice along the path $\Gamma-K-K'-\Gamma$ under various TRS-breaking effects, with the Chern numbers labeled on each nondegenerate band: (a) $\Omega=0$, (b) $\Omega=0.3$, (c) $\Omega=\Omega_0\approx0.602$ and (d) $\Omega=0.7$, and topologically nontrivial gaps are filled with yellow shade. From the uppermost to the lowermost, the 4 bands are respectively denoted with TO (transverse optical), LO (longitudinal optical), LA (longitudinal acoustic) and TA (transverse acoustic). (e)-(h) The phononic dispersion of a gyroscope-coupled graphene nanoribbon with 12 zigzag chains, the leftmost and rightmost of which are fixed, under various TRS-breaking field strengths: (e) $\Omega=0$, (f) $\Omega=0.3$, (g) $\Omega=\Omega_0\approx0.602$ and (h) $\Omega=0.7$. The black curves are bulk bands, and the red/blue/green curves correspond to edge states localized at left/right/both side(s) (we define a vibrational mode to be an edge state if the two side chains contribute more than 50\% of average kenetic energy).}
    \label{dispersion}
\end{figure*}

Previous studies have shown phononic topology of honeycomb lattice under different TRS-breaking mechanisms~\cite{Kariyado2015,Lifa2010,YT2015}. Here we also apply our theory of gyroscope-coupled lattice vibration to a honeycomb lattice as shown in Fig. \ref{structure}(a) and (b). Let $a$ be the NN distance and $c_1$, $c_2$ denote the spring stiffness of NN and NNN couplings, respectively. Without loss of generality we set the parameters $a=M=c_1=1$ and $c_2=g=0.05$. We solve for the phononic dispersion under various TRS-breaking field strengths and calculate the Chern numbers of each nondegenerate band.

Figure \ref{dispersion}(a)-(d) shows typical phononic dispersions, the corresponding Chern numbers and topological nature of band gaps. When TRS is preserved (i.e. $\Omega=0$), the four branches are connected at either $\Gamma$ or $K$ and $K^\prime$ thus form a gapless band structure. As a finite $\Omega$ is turned on, the four bands are never connected with each other, forming 3 finite band gaps. When $\Omega$ reaches a critical value $\displaystyle\Omega_0 = \frac{3c_1/(4M)}{\sqrt{3c_1/(2M)+g}} \approx 0.602$, the LA and LO modes touch at $\Gamma$ thus close the band gap between LA and LO. When $\Omega>\Omega_0$ the gap between LA and LO reopens but its gap Chern number is changed to $0$. All above are qualitatively precedent in previous study in Coriolis field~\cite{YL2017,YT2015}, except for one difference: at $\Gamma$ the acoustic bands do not start from 0, because the gravity of gyroscopes breaks the continuous translational symmetry (or the acoustic sum rule), i.e. an overall displacement of the whole lattice does lead to an increase in potential energy.

As a direct result of non-zero phononic Chern numbers, the honeycomb lattice is expected to support one-way edge states. To observe the edge states, we calculate the phonon dispersion of zigzag honeycomb ribbons with 12 zigzag chains and fixed boundary condition [i.e. the zigzag chains of the two edges are fixed] under different $\Omega$'s [Fig. \ref{dispersion}(e)-(h)]. When TRS is preserved, the finite size effect may open a small band gap at the Dirac points. However the phonon dispersion is gapless due to the existence of flat edge states (green lines) which is protected by the $\pi$ Berry phase of each Dirac cone~\cite{Delplace2011,SM,Zak1989}. When a small $\Omega$ is tuned on, the bulk phonon spectrum become gapped but with gapless chiral (one-way) edge states connecting different bulk bands. As $\Omega$ increases, the edge states connecting bulk LA and LO bands disappear after $\Omega$ exceeds $\Omega_0$. All these are consistent with the Chern number calculation and previous works~\cite{YT2015}.

\section{IV. Nonlinear effect on one-way edge states}

Previous studies have shown that the topological edge states are robust against backscatterings induced by lattice boundary imperfection \cite{P2015} or disorder \cite{YL2017}. Such backscattering mechanisms scatter phonon by reflection which preserves the frequency of phonon. However, in phononic systems, there is another important effect---nonlinear effect, also known as anharmonic effect---which leads to phonon-phonon interactions. Whether the topological chiral phonon states remains stable under nonlinear effects is still an open question, which is essential in determining the phonon transporting ability.

\subsection{A. Method}

We construct a finite honeycomb lattice with fixed boundary, $n=296$ moveable sites [Fig. \ref{spec}(a)] and selected parameters $a=c_1=1$, $c_2=0$, $h=0.5$, $I_0=0.25$ ($M=1$), $g=0.9$ and $\Omega=0.3$. By solving the linear approximate dynamic equation we obtain monochromatic vibrational modes of the form $\bm{U}=\bm{U}_0{\rm e}^{{\rm i}\omega_0t}$, where $\bm{U}_0=(\bm{U}_0^1,\ldots,\bm{U}_0^n)$ is normalized, and $\bm{U}_0^j=(U_0^{jx},U_0^{jy})$ stands for the in-plane vibrational complex amplitude vector of the $j^{\text{th}}$ site. Due to finite lattice effect the spectrum of $\omega_0$ is actually discrete, and naturally those values within nontrivial bulk band gaps correspond to vibrational modes confined to the edge, which we call ``edge states''.

We study the nonlinear scattering process of 4 edge states with $\omega_0=1.00535$, 1.10228, 1.20023 and 2.00888 by molecular dynamics simulation. The first three lie in the TA/LA gap and the last one is within LO/TO gap [Fig. \ref{spec}(e)]. To study the edge state with angular frequency $\omega_0$, we note that for the harmonic oscillation $\bm{U}=\bm{U}_0{\rm e}^{{\rm i}\omega_0t}$, ${\rm Re}(\bm{U}_0)$ is the initial displacement and $\omega_0{\rm Im}(\bm{U}_0)$ is the initial velocity, and in our finite lattice we set the initial conditions to be
\begin{equation}
    \left\{\begin{array}{>{\displaystyle}l}
        \left.\bm{x}_j\right|_{t=0}=\frac{0.6}{\omega_0}{\rm Re}(\bm{U}_0^j)\\
        \left.\left(z_j^2\frac{\rm d}{{\rm d}t}\frac{\bm{x}_j}{z_j}\right)\right|_{t=0}=0.6{\rm Im}\left(\bm{U}_0^j\right)
    \end{array}\right.
\end{equation}
where $\bm{x}_j=\bm{U}_j/h=(x_j,y_j)$, $\bm{U}_j$ is the in-plane displacement of the $j^{\text{th}}$ site, and $z_j=\sqrt{1-x_j^2-y_j^2}$. This initial condition guarantees that the average vibrational kinetic energy of the state is fixed under linear approximation. Under this initial condition, we solve Eq. (\ref{strictdyn}) numerically for the time evolution of the state from $t=0$ to $t=2400\pi$ ($\approx7540$) (see the detailed method in SM~\cite{SM}).

In order to view the change of vibrational frequency during the scattering process, we need to analyze the frequency spectrum of vibration by Fourier analysis. We define the spectrum of the $j^{\text{th}}$ site around $t=t_0$ to be
\begin{equation}
	\bm{X}_j(\omega;t_0)=\frac{1}{\Delta T}\int_{t_0-\frac{\Delta T}{2}}^{t_0+\frac{\Delta T}{2}}\bm{x}_j(t){\rm e}^{-{\rm i}\omega t}{\rm d}t
    \label{Fourier}
\end{equation}
where we set the Fourier analysis range to be $\Delta T=160\pi$ so that the frequency resolution is ${\rm d}\omega=2\pi/\Delta T=0.0125$ (we also tried the case of $\Delta T=80\pi$, $240\pi$, $320\pi$ and $400\pi$, and there is no qualitative difference). Also note from the phononic dispersion [Fig. \ref{spec}(e)] that the maximum vibrational frequency we need to observe is below $\omega_m=2.5$, according to Nyquist-Shannon sampling theorem, a discretization by time interval ${\rm d}t<\pi/\omega_m=0.4\pi$ allows for reconstruction of the whole frequency spectrum. Thus in Eq. (\ref{Fourier}) we use summation to approximate the integration by setting ${\rm d}t=0.2\pi$.

Moreover, we select $n_e=98$ sites near the boundary, whose indexes form \textit{edge set}, to represent edge vibration. We also select $n_b=70$ sites in the bulk, whose indexes form \textit{bulk set}, to represent bulk vibration, as indicated in Fig. \ref{spec}(a). We define the \textit{edge spectrum} around $t_0$ to be
\begin{equation}
	X_e(\omega;t_0)=\sqrt{\frac{1}{n_e}\sum_{j\in\text{edge set}}\left|\bm{X}_j(\omega;t_0)\right|^2}
\end{equation}
and define the \textit{bulk spectrum} similarly.

\subsection{B. Results}

\begin{figure*}[t]
    \centering
    \includegraphics[width=0.96\textwidth,angle=0]{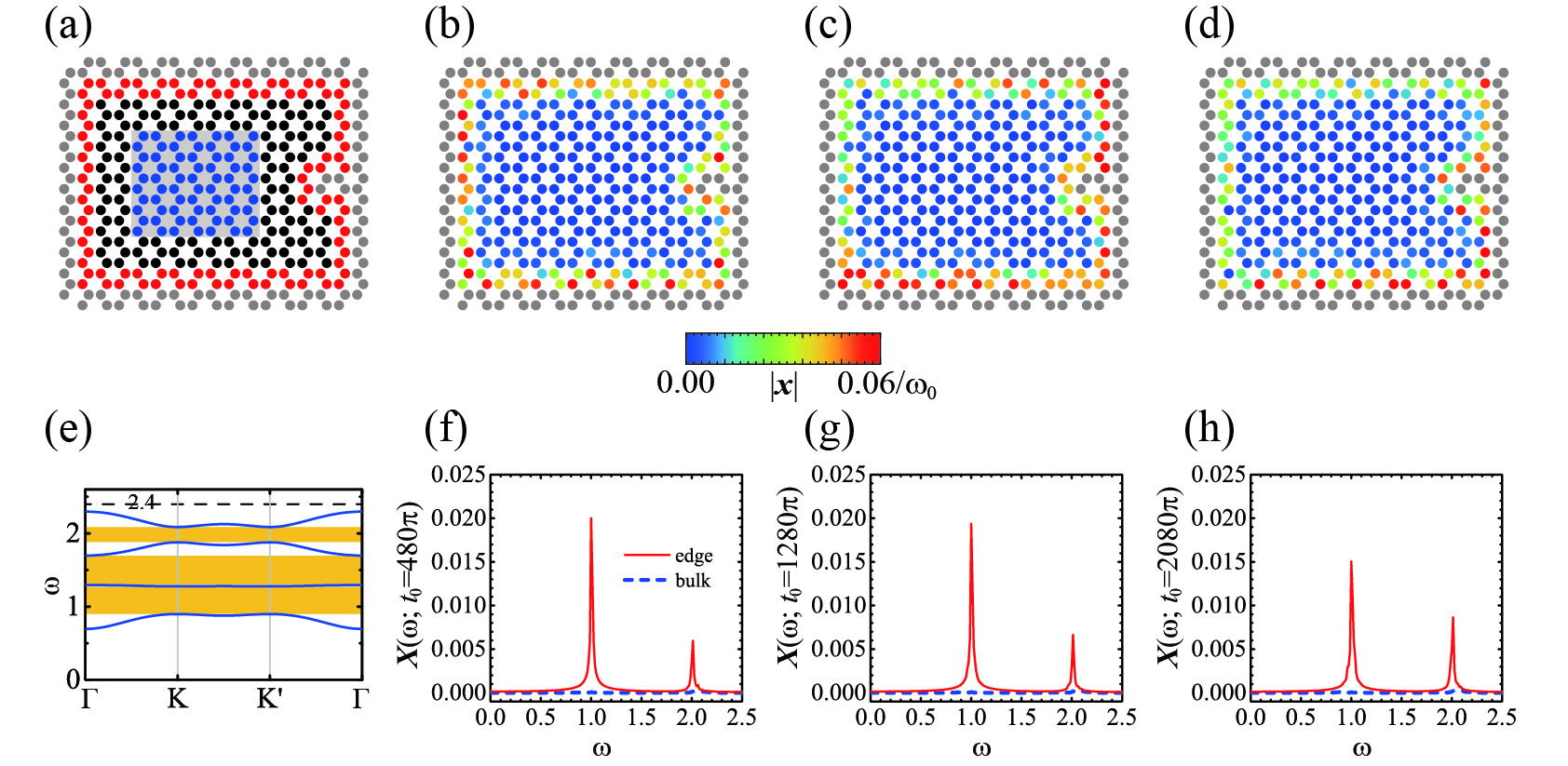}
    \caption{(a) A finite honeycomb lattice with 404 sites. The 108 gray circles represent fixed sites, serving as fixed boundary and edge defect, and all the others are moveable. The red circles (blue circles in the gray rectangular shade) are chosen to represent edge (bulk) vibration. (b)-(d) Snapshots of evolution of the edge state with $\omega_0=1.00535$ at (b) $t=480\pi$, (c) $t=1280\pi$ and (d) $t=2080\pi$. (e) The bulk phononic dispersion under our chosen parameters, with the nontrivial gaps shaded in yellow. (f)-(h) The edge and bulk vibration spectra of evolution of the edge state with $\omega_0=1.00535$ around (f) $t_0=480\pi$, (g) $t_0=1280\pi$ and (h) $t_0=2080\pi$.}
    \label{spec}
\end{figure*}

Figure \ref{spec}(b)-(d) show snapshots of evolution of the edge state $\omega_0=1.00535$ at $t=480\pi$, $1280\pi$ and $2080\pi$. For other edge states please see SM~\cite{SM}. The vibrations of edge states with $\omega_0=1.00535$, 1.20023 and 2.00888 mostly keep confined at the edge, yet that with $\omega_0=1.1$ dissipates a lot into the bulk.

\begin{figure*}[t]
    \centering
    \includegraphics[width=0.96\textwidth,angle=0]{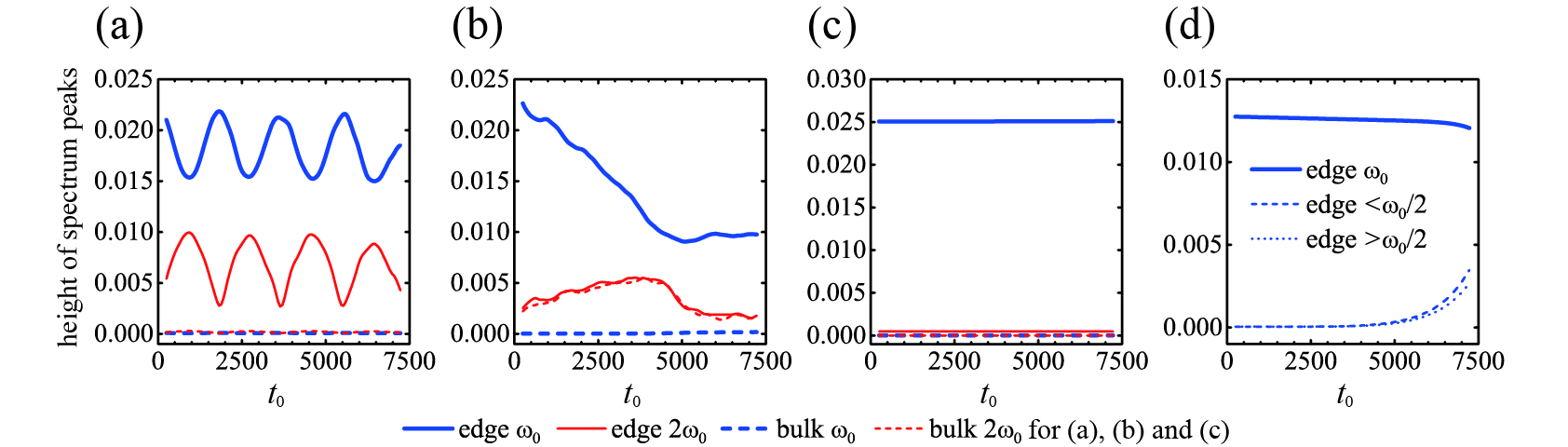}
    \caption{The time evolution of spectrum peaks in the systems under monochromatic edge inputs: (a) $\omega_0=1.00535$, (b) $\omega_0=1.10228$, (c) $\omega_0=1.20023$ and (d) $\omega_0=2.00888$.}
    \label{peakev}
\end{figure*}

Edge and bulk spectra for evolution of the edge state $\omega_0=1.00535$ around $t_0=480\pi$, $1280\pi$ and $2080\pi$ are presented in Fig. \ref{spec}(f)-(h), and those for other edge states are presented in the SM~\cite{SM}. For evolution of the edge state $\omega_0=1.00535$, we see two peaks at $\omega=\omega_0$ and $\omega=2\omega_0$ in edge spectra and no peak in bulk spectra. For $\omega_0=1.10228$, we first see two peaks at $\omega_0$ and $2\omega_0$ in the edge spectrum and one peak at $2\omega_0$ in the bulk spectrum. Then as $t_0$ increases, peaks at $\omega\approx0.9$, $\omega\approx1.3$, $\omega\approx1.8$ and $\omega\approx2.0$ emerge. For $\omega_0=1.20023$, we only see one peak at $\omega_0$ and a very small peak at $2\omega_0$ in edge spectra and nothing in bulk spectra. For $\omega_0=2.00888$, most of the time there is only one peak at $\omega_0$ in the edge spectrum and a very small peak at $\omega_0$ in the bulk spectrum. After a relatively long time, several small peaks around $\omega_0/2$ emerge, of which the nearest two to $\omega_0/2$ play the major role.

Furthermore, we plot the time evolution of height of the peaks at $\omega_0$ and $2\omega_0$ for the three TA/LA edge states in Fig. \ref{peakev}(a)-(c), and plot those of peaks at $\omega_0$ and nearest to $\omega_0/2$ in Fig. \ref{peakev}(d). We see a periodical variation in level of $\omega_0=1.00535$ with period slightly less than 2000, an unsteady decline in level of $\omega_0=1.10228$, a nearly constant level of $\omega_0=1.20023$ and a steady level of $\omega_0=2.00888$ that can maintain for a finite time.

\subsection{C. Discussion}

The nonlinear part of lattice dynamic equations are well-known to give rise to phonon-phonon scattering process. With relatively small vibrational amplitude, 3-phonon processes play the major role of the scattering effect. For a monochromatic input with angular frequency $\omega_0$, there can be two kinds of primary 3-phonon processes: collision between two phonons, which gives
$$
    \omega_0+\omega_0\rightarrow2\omega_0
$$
and splitting of a single phonon, which makes two phonons, one of which must have its angular frequency no larger than $\omega_0/2$. Further processes may include further collision like $2\omega_0+\omega_0\rightarrow3\omega_0$, and splitting of $2\omega_0$, etc. Some of these processes may be prohibited due to restriction on vibrational frequency by phononic dispersion. For the 4 inputs investigated in this study, all 3-phonon scattering processes that are \textit{theoretically} allowed by the phononic dispersion are indicated in Table \ref{table1}.

\begin{table}
\caption{Theoretically allowed 3-phonon processes of monochromatic input $\omega_0$, indicated by $\surd$.}
\begin{tabular}{|c|c|c|c|c|}
	\hline
    $\omega_0$ & 1.00535 & 1.10228 & 1.20023 & 2.00888 \\
    \hline
    $\omega_0+\omega_0\rightarrow2\omega_0$ & $\surd$ & $\surd$ & & \\\hline
    $2\omega_0+\omega_0\rightarrow3\omega_0$ & & & & \\\hline
    splitting of $\omega_0$ & & & & $\surd$ \\\hline
    splitting of $2\omega_0$ & $\surd$ & $\surd$ & & \\\hline
    \parbox[c]{70pt}{\centering$\omega_0$ collides with splitted phonons} & $\surd$ & $\surd$ & & \\\hline
    \parbox[c]{70pt}{\centering$2\omega_0$ collides with splitted phonons} & & & & \\\hline
\end{tabular}
\label{table1}
\end{table}

Now we focus on the case of $\omega_0=2.00888$. Though the splitting channel is theoretically allowed, the observed result is that the edge state splits mainly into other edge states rather than bulk states. To understand this, we note that the overlap between wavefunctions of edge states and bulk states, as well as those between edge states and edge states, is confined to edge. Due to normalization of wavefunctions, we can expect the ``scattering matrix element'' between edge and bulk states to be much smaller than that between edge and edge states. Another feature we see is that the splitting process takes a larger time scale than colliding processes, but occurs much more abruptly.

For the other three cases $\omega_0=1.00535$, 1.10228 and 1.20023, the splitting channel is prohibited, thus we consider the collision which makes $2\omega_0$. For $\omega_0=1.20023$, $2\omega_0$ is beyond all the bands, thus collision is also prohibited and no 3-phonon process occurs, which is in consistent with our observation of constant level of $\omega_0$ vibration. The tiny peak at $2\omega_0$ is due to the fact that the vibration is not exactly sinusoidal in anharmonic potential, which leads to high-frequency components. As for $\omega_0=1.10228$, $2\omega_0$ is within the TO band thus is a bulk state, accounting for our observation that the edge state dissipates into the bulk. This bulk state, in turn, splits into other states which also include bulk ones, and then further collision occurs, etc, giving all the peaks we see in the spectrum.

For $\omega_0=1.00535$, $2\omega_0$ lies in the LO/TO gap thus is still confined to the edge. According to our previous argument further splitting process of $2\omega_0$ will also create edge states. This is indeed the case according to our simulation, since the fluctuation can be explained as oscillation of the reversible process $\omega_0+\omega_0\leftrightarrow2\omega_0$. What is different is that the time scale for the splitting process is much smaller. A presumable inference is that the existence of $\omega_0$ phonons can help to stimulate the splitting of $2\omega_0$, in analogy to the induced radiation of atoms. To understand from an alternative perspective, due to the ``particle-hole'' symmetry, the existence of $\omega_0$ phonons indicates the existence of ``holes'' with angular frequency $-\omega_0$, which collide with $2\omega_0$ phonons making $\omega_0$ phonons. In addition, the periodicity may be explained by a long-period variation of coupling coefficient, whose period can be estimated by $\pi/(2\times1.00535-2.00888)\approx1726$ (the anharmonic perturbation on vibrational frequency may affect this period significantly, depending on the amplitude). In other words, the periodicity may result from detuning between $2\omega_0$ and 2.00888, brought about by finite lattice effect. In an infinitely long ribbon, since frequency distribution of edge modes is continuous, an edge state with angular frequency $\omega_0$ may couple with a range of states around $2\omega_0$ by different detuning effects. The range in frequency gives rise to finite phonon lifetime.

In a general TRS-broken nonlinear phononic crystal, the possible behaviors of a small-amplitude topological edge state with angular frequency $\omega_0$ fall into 3 cases:

(i) If $2\omega_0$ is within a topologically trivial gap (not considered in our finite lattice model) or outside any bulk band, and in the meanwhile no nontrivial gap is below $2\omega_0$, then collision is prohibited and splitting is also suppressed. In this case the edge state is the most robust against nonlinear scattering, thus is a desirable candidate for application in energy conduction and signal transmission in phonon devices. E.g. the edge state with $\omega_0=1.20023$ in our finite system.

(ii) If $2\omega_0$ is within a nontrivial gap and/or there is nontrivial gaps below $\omega_0$ allowing for splitting into edge states, with the same sign of gap Chern numbers, then the edge state is not robust but still keeps confined to the edge and maintains its transport direction. Such edge states, with near-zero dissipation, may also be used for energy transport. E.g. $\omega_0=1.00535$ and 2.00888 in our system.

(iii) If $2\omega_0$ is within a bulk band, then the state is not robust and easily dissipates into the bulk. E.g. $\omega_0=1.10228$ in our system, as observed in the molecular dynamics simulation.

What we need to point out is that robustness against or sensitiveness to nonlinear scattering is not a result of the state's topological nature, unlike robustness against defect or disorder. Actually any state will evolve to thermal equilibrium after sufficiently long time. The difference is only a matter of time scale needed for the scattering or dissipation process. The direct determinant is the phononic dispersion, as we have seen from the above argument. However, by tuning the parameters of lattice, like the spring coupling constant, etc, we can change the phononic dispersion to make certain edge states satisfy the robust or dissipationless conditions discussed above in (i) and (ii), realizing selective transport of edge states. While this may be difficult to realize in natural materials, this can be done in artificially designed metamaterials.

Low-dissipation transport of nonlinear topological phononic edge states may find their importance in future devices like low-power one-way phonon waveguides and phonon diodes~\cite{YL2017}. There are mainly two required conditions for such phonon devices to work with near-zero resistance:

\paragraph{i. Relatively small vibrational amplitude}If the amplitude is large, then the 4-phonon processes and those involving more phonons will become as significant as 3-phonon processes. Those multi-phonon processes are much more likely to produce high diversity of edge and bulk phonons.

\paragraph{ii. Low temperature}If the temperature is not low enough, then there are lots of background phonons in thermal equilibrium, ranging all over the bands, which can scatter off the edge states, leading to a thermal resistance in analogy to electric resistance in metals.

\section{V. Conclusion}

We have demonstrated that the TRS-breaking term of gyroscopic phononic crystals is equivalent to a Coriolis field, solving a contradiction left by Refs.~\cite{P2015,CARTA2014,Brunrspa2012} and confirming the universality of Ref.~\cite{YL2017}'s general argument on TRS-breaking terms. Furthermore, we have revealed the unrobust nature of one-way topological phononic edge states against nonlinear phonon-phonon scattering effect and specified some conditions under which some edge states are more robust or unlikely to be dissipated into bulk, as tested by the molecular dynamic simulation. This inspires an idea of suppressing the scattering process while designing the materials to be used for phonon devices.

Most importantly, our work may open up a new researching subject dealing with nonlinear effect on topological phononic states. This work only provides very basic ideas and a lot of further theorizations, calculations and simulations are still required, such as to calculate the phonon lifetime, to predict the phononic resistance-temperature characteristic of edge states, or whether edge states can be localized to solitons, etc.

\begin{acknowledgments}

J.S. acknowledges support from his supervisor and fellow in State Key Laboratory of Low Dimensional Quantum Physics.

\end{acknowledgments}

\bibliography{bibliography}

\renewcommand\thefigure{S\arabic{figure}}

\begin{widetext}

\begin{center}
\textbf{Supplementary materials to \textit{One-way topological edge states in nonlinear gyroscopic phononic crystals}}
\end{center}

\section{I. Exact dynamic equations and method for numerical solution}

Now that in the main text we have obtained the expression for the moment $\bm{N}$ under $O-x'y'z'$ coordinates:
\begin{equation}
    \left\{\begin{array}{>{\displaystyle}l}
        N_{x'}=I_0(\ddot{\theta}-\dot{\phi}^2\sin{\theta}\cos{\theta})+I_3\Psi\dot{\phi}\sin\theta\\
        N_{y'}=I_0(\ddot{\phi}\sin\theta+2\dot{\phi}\dot{\theta}\cos\theta)-I_3\Psi\dot{\theta}\\
        N_{z'}=0
    \end{array}\right.
\end{equation}
we transform it into $O-xyz$ coordinates:
\begin{equation}
    \left\{\begin{array}{>{\displaystyle}l}
        N_x=I_0(\ddot{\theta}\cos\phi-\dot{\phi}^2\cos\phi\sin\theta\cos\theta-\ddot{\phi}\sin\phi\sin\theta\cos\theta-2\dot{\phi}\dot{\theta}\sin\phi\cos^2\theta)+I_3\Psi(\dot{\phi}\cos\phi\sin\theta+\dot{\theta}\sin\phi\cos\theta)\\
        N_y=I_0(\ddot{\theta}\sin\phi-\dot{\phi}^2\sin\phi\sin\theta\cos\theta+\ddot{\phi}\cos\phi\sin\theta\cos\theta+2\dot{\phi}\dot{\theta}\cos\phi\cos^2\theta)+I_3\Psi(\dot{\phi}\sin\phi\sin\theta-\dot{\theta}\cos\phi\cos\theta)
    \end{array}\right.
    \label{myN_S}
\end{equation}
According to our definition of the ``reduced displacement'' $x=U_x/h=\sin\phi\sin\theta$ and $y=U_y/h=-\cos\phi\sin\theta$, it can be checked by some calculation that Eq. (\ref{myN_S}) is equivalent to the following expressions (here we assume $\theta<\pi/2$ so that $\sqrt{1-x^2-y^2}=\cos\theta$):
\begin{equation}
    \left\{\begin{array}{>{\displaystyle}l}
        N_x=-I_0\frac{\rm d}{{\rm d}t}\left(\left(1-x^2-y^2\right)\frac{\rm d}{{\rm d}t}\frac{y}{\sqrt{1-x^2-y^2}}\right)+I_3\Psi\dot{x}\\
        N_y=I_0\frac{\rm d}{{\rm d}t}\left(\left(1-x^2-y^2\right)\frac{\rm d}{{\rm d}t}\frac{x}{\sqrt{1-x^2-y^2}}\right)+I_3\Psi\dot{y}
    \end{array}\right.
    \label{myNxy}
\end{equation}
From the force analysis $\bm{N}=\bm{h}\times\bm{T}+\bm{l}\times\bm{G}$ we have:
\begin{equation}
    \left\{\begin{array}{>{\displaystyle}l}
        N_x=h\left(yT_z-\sqrt{1-x^2-y^2}T_y\right)+lGy\\
        N_y=h\left(\sqrt{1-x^2-y^2}T_x-xT_z\right)-lGx
    \end{array}\right.
    \label{forceNxy}
\end{equation}
Combining Eqs. (\ref{myNxy}) and (\ref{forceNxy}), we get the exact dynamic equations for an arbitrary site labelled by the $i$th ($1\leq i\leq n$, $n$ is the total number of moveable sites):
\begin{equation}
    \left\{\begin{array}{>{\displaystyle}l}
        I_0\frac{\rm d}{{\rm d}t}\left(\left(1-x_i^2-y_i^2\right)\frac{\rm d}{{\rm d}t}\frac{x_i}{\sqrt{1-x_i^2-y_i^2}}\right)+I_3\Psi\dot{y}_i+h\left(x_iT_z^i(x_k,y_k)-\sqrt{1-x_i^2-y_i^2}T_x^i(x_k,y_k)\right)+lGx_i=0\\
        I_0\frac{\rm d}{{\rm d}t}\left(\left(1-x_i^2-y_i^2\right)\frac{\rm d}{{\rm d}t}\frac{y_i}{\sqrt{1-x_i^2-y_i^2}}\right)-I_3\Psi\dot{x}_i+h\left(y_iT_z^i(x_k,y_k)-\sqrt{1-x_i^2-y_i^2}T_y^i(x_k,y_k)\right)+lGy_i=0
    \end{array}\right.
    \label{stri}
\end{equation}
where $(x_k,y_k)$ stands for $(x_1,y_1,\ldots,x_n,y_n)$.

To solve the equations numerically, we first transform the second-order differential equations equivalently to first-order ones by introducing $2n$ auxiliary variables, $\xi_i$ and $\eta_i$ ($1\leq i\leq n$):
\begin{equation}
	\left\{\begin{array}{>{\displaystyle}l}
		\xi_i=(1-x_i^2-y_i^2)\frac{\rm d}{{\rm d}t}\frac{x_i}{\sqrt{1-x_i^2-y_i^2}}\\
        \eta_i=(1-x_i^2-y_i^2)\frac{\rm d}{{\rm d}t}\frac{y_i}{\sqrt{1-x_i^2-y_i^2}}
    \end{array}\right.
    \label{aux_var}
\end{equation}
From Eq. (\ref{aux_var}) we can solve for $\dot{x}_i$ and $\dot{y}_i$:
\begin{equation}
	\left\{\begin{array}{>{\displaystyle}l}
		\dot{x}_i=\frac{(1-x_i^2)\xi_i-x_iy_i\eta_i}{\sqrt{1-x_i^2-y_i^2}}\\
        \dot{y}_i=\frac{(1-y_i^2)\eta_i-x_iy_i\xi_i}{\sqrt{1-x_i^2-y_i^2}}
    \end{array}\right.
    \label{xyderiv}
\end{equation}
A substitution of Eqs. (\ref{aux_var}) and (\ref{xyderiv}) into Eq. (\ref{stri}) yields
\begin{equation}
	\left\{\begin{array}{>{\displaystyle}l}
        I_0\dot{\xi}_i+I_3\Psi\frac{(1-y_i^2)\eta_i-x_iy_i\xi_i}{\sqrt{1-x_i^2-y_i^2}}+lWx_i+hT_z^i(x_k,y_k)x_i-h\sqrt{1-x_i^2-y_i^2}T_x^i(x_k,y_k)=0\\
        I_0\dot{\eta}_i-I_3\Psi\frac{(1-x_i^2)\xi_i-x_iy_i\eta_i}{\sqrt{1-x_i^2-y_i^2}}+lWy_i+hT_z^i(x_k,y_k)y_i-h\sqrt{1-x_i^2-y_i^2}T_y^i(x_k,y_k)=0
    \end{array}\right.
    \label{xietaderiv}
\end{equation}
Equations (\ref{xyderiv}) and (\ref{xietaderiv}) together give the first-order version of dynamic equations, which is technologically no difficult for computers to solve numerically. Here for the study of vibrational mode $\bm{U}=\bm{U}_0{\rm e^{{\rm i}\omega_0t}}$, the initial conditions are set by
\begin{equation}
    \left\{\begin{array}{>{\displaystyle}l}
        \left.(x_i,y_i)\right|_{t=0}=\frac{0.6}{\omega_0}{\rm Re}(\bm{U}_0^i)\\
        \left.(\xi_i,\eta_i)\right|_{t=0}=0.6{\rm Im}(\bm{U}_0^i)
    \end{array}\right.
\end{equation}

\section{II. Introduction to phononic topology}

Now that we have the Schr\"{o}dinger-like equation~\cite{YL2017} $H(\bm{k})\psi_{n\bm{k}}=\omega_n(\bm{k})\psi_{n\bm{k}}$ for the $n$th band, we define the Berry connection (assume normalized $\psi_{n\bm{k}}$)
\begin{equation}
    \bm{A}_n(\bm{k})={\rm i}\psi_{n\bm{k}}^\dag\nabla_{\bm{k}}\psi_{n\bm{k}}
\end{equation}
and the Berry curvature
\begin{equation}
	B_n(\bm{k})=\nabla_{\bm{k}}\times\bm{A}_n={\rm i}\nabla_{\bm{k}}\psi_{n\bm{k}}^\dag\times\nabla_{\bm{k}}\psi_{n\bm{k}}
\end{equation}
The gauge-invariant calculation formula for Berry curvature is
\begin{equation}
	B_n(\bm{k})={\rm i}\sum_{m\neq n}\frac{\psi_{n\bm{k}}^\dag(\nabla_{\bm{k}}H)\psi_{m\bm{k}}\times\psi_{m\bm{k}}^\dag(\nabla_{\bm{k}}H)\psi_{n\bm{k}}}{\left(\omega_n(\bm{k})-\omega_m(\bm{k})\right)^2}
\end{equation}
where the summation is over all positive and negative bands other than the $n$th. The Chern number of the $n$th band of a 2-dimensional lattice is defined as
\begin{equation}
    \mathcal{C}_n=\frac{1}{2\pi}\int_{\rm BZ}B_n(\bm{k}){\rm d}^2\bm{k}
\end{equation}
Since it can be proven that $\mathcal{C}_n$ is always an integer, the Chern number describes an important topological property of the band. In a finite lattice system, nontrivial topological properties can lead to one-way edge vibrational modes within bulk gaps, whose chirality is determined by the sign of gap Chern number. If the gap Chern number is zero, then no edge state is in the gap.

Under spatial inversion symmetry (SIS), we have $\omega_n(-\bm{k})=\omega_n(\bm{k})$ and $B_n(-\bm{k})=B_n(\bm{k})$; under time-reversal symmetry (TRS), $\bm{\eta(k)}=0$, $\omega_n(-\bm{k})=\omega_n(\bm{k})$, $B_n(-\bm{k})=-B_n(\bm{k})$ and all the Chern numbers vanish. Thus in order to realize nonzero Chern number, we must break TRS; if we want nonzero Berry curvature, we need to break either SIS or TRS.

The quantization of Berry phase over 1D BZ (known as Zak phase~\cite{Zak1989}) under inversion symmetry in 1D lattice also has profound topological significance. P. Delplace et al~\cite{Delplace2011} generalized Zak phase to 2D graphene lattice, in which they focused on the Berry phase along a certain direction in BZ over one period and showed a relationship between Zak phase and the existence of edge states in nanoribbons without TRS breaking. Here we give a brief talk on the adoption of Zak phase theory into our phononic system, which may help account for the flat edge states we find between LO and LA bands under preserved TRS.

As one can easily show that under preserved TRS the Berry connection has a simple form of $\bm{A}_n(\bm{k})={\rm i}\bm{u}_{n\bm{k}}^\dag\nabla_{\bm{k}}\bm{u}_{n\bm{k}}$ (assume $\bm{u}_{n\bm{k}}=(u_{\bm{k},Ax},u_{\bm{k},Ay},u_{\bm{k},Bx},u_{\bm{k},By})^T$ to be normalized), by definition the Zak phase along $x$ direction as a function of $k_y$ is
\begin{equation}
    \mathcal{Z}_n(k_y)={\rm i}\int_{\frac{G_x}{2}}^{\frac{G_x}{2}}\bm{u}_{n\bm{k}}^\dag\frac{\partial\bm{u}_{n\bm{k}}}{\partial k_x}{\rm d}k_x
\end{equation}
where $G_x$ is the total width of 1st BZ in $x$ direction. Note that the system displays mirror symmetry about $y$-axis, which is the generalization of 1D inversion symmetry, it can be shown that $\mathcal{Z}_n(k_y)$ must be an integer multiple of $\pi$ (assume $x$-direction periodicity in $\bm{k}$-space i.e. $\bm{u}_{n(k_x+G_x,k_y)}=\bm{u}_{n(k_x,k_y)}$). Moreover, any gauge transform $\bm{u}_{n\bm{k}}\rightarrow{\rm e}^{{\rm i}\zeta_n(\bm{k})}\bm{u}_{n\bm{k}}$ that respects the $x$-direction $\bm{k}$-space periodicity only changes the Zak phase by multiples of $2\pi$. Therefore, the remainder of $\mathcal{Z}_n(k_y)$ modulo $2\pi$ is a gauge-invariant topological number. In a nanoribbon with large but finite size in $x$-direction, $\mathcal{Z}_n(k_y)\equiv\pi$ indicates edge states corresponding to the along-edge wavevector $k_y$ while $\mathcal{Z}_n(k_y)\equiv0$ indicates no edge states. Since the singularity at each Dirac point contributes a Berry phase of $\pi$, a topological transition occurs when $k_y$ is swept over the $y$-axis projection of a Dirac point. Specifically $\mathcal{Z}_n(k_y)$ modulo $2\pi$ changes from $\pi$ to 0 when $k_y$ increases and exceeds $K_y$ (the $y$-coordinate of the Dirac point $\bm{K}$). This is why in $k_y$-space all the flat edge states are found between $K_y$ and $K'_y$.

When TRS is broken, the mirror symmetry is also broken because under mirror operation the TRS-breaking field changes its sign. In this case Zak phase is no longer quantized and the edge states become associated with Chern number rather than Zak phase.

\section{III. Additional results for simulation and spectrum analysis}

\begin{figure}[ht]
    \centering
    \includegraphics[width=0.96\textwidth,angle=0]{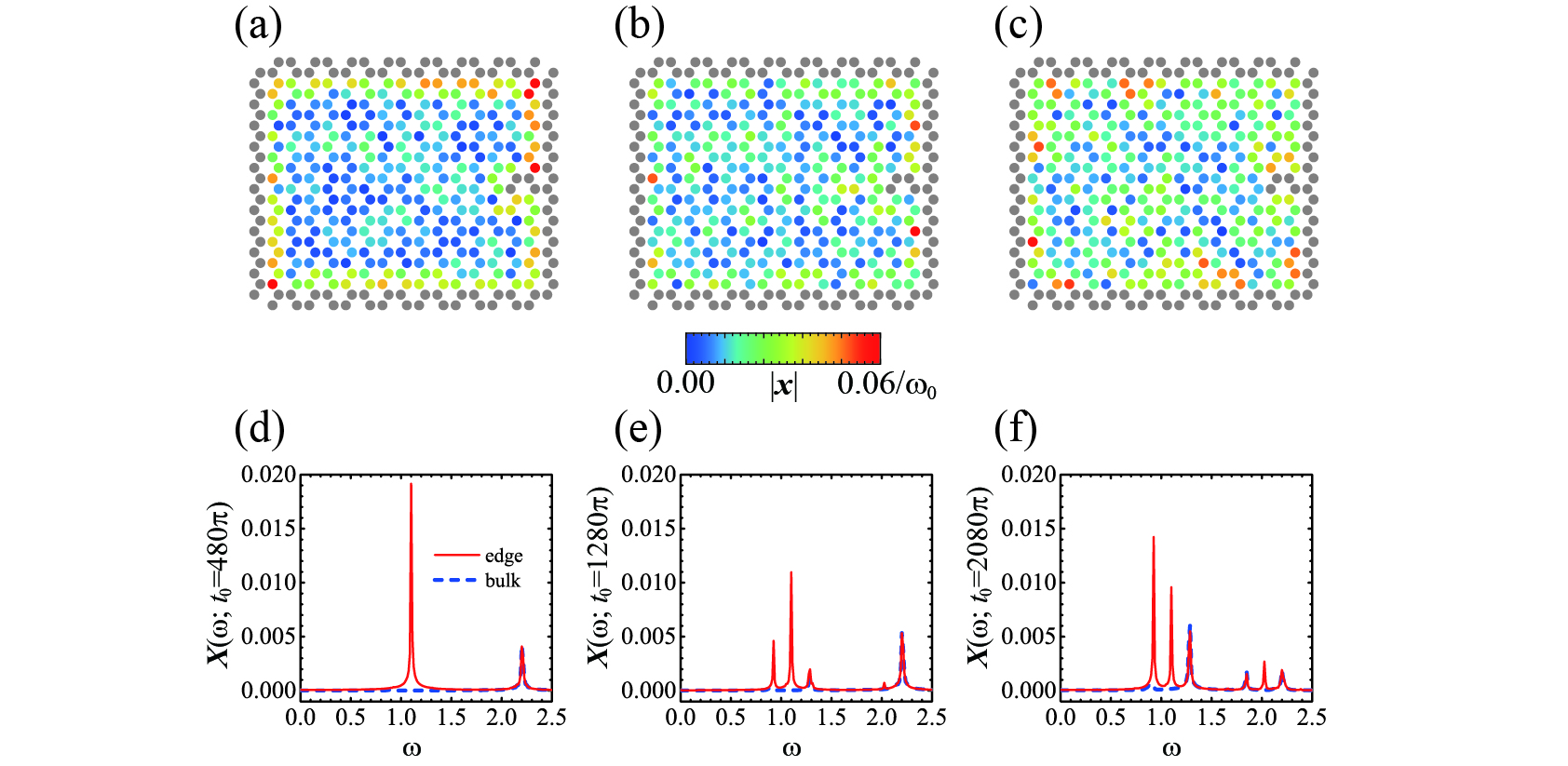}
    \caption{(a)-(c) Snapshots of evolution of the edge state with $\omega_0=1.10228$ at (a) $t=480\pi$, (b) $t=1280\pi$ and (c) $t=2080\pi$. (d)-(f) The edge and bulk vibration spectra of evolution of the edge state with $\omega_0=1.10228$ around (d) $t_0=480\pi$, (e) $t_0=1280\pi$ and (f) $t_0=2080\pi$.}
\end{figure}
\begin{figure}[ht]
    \centering
    \includegraphics[width=0.96\textwidth,angle=0]{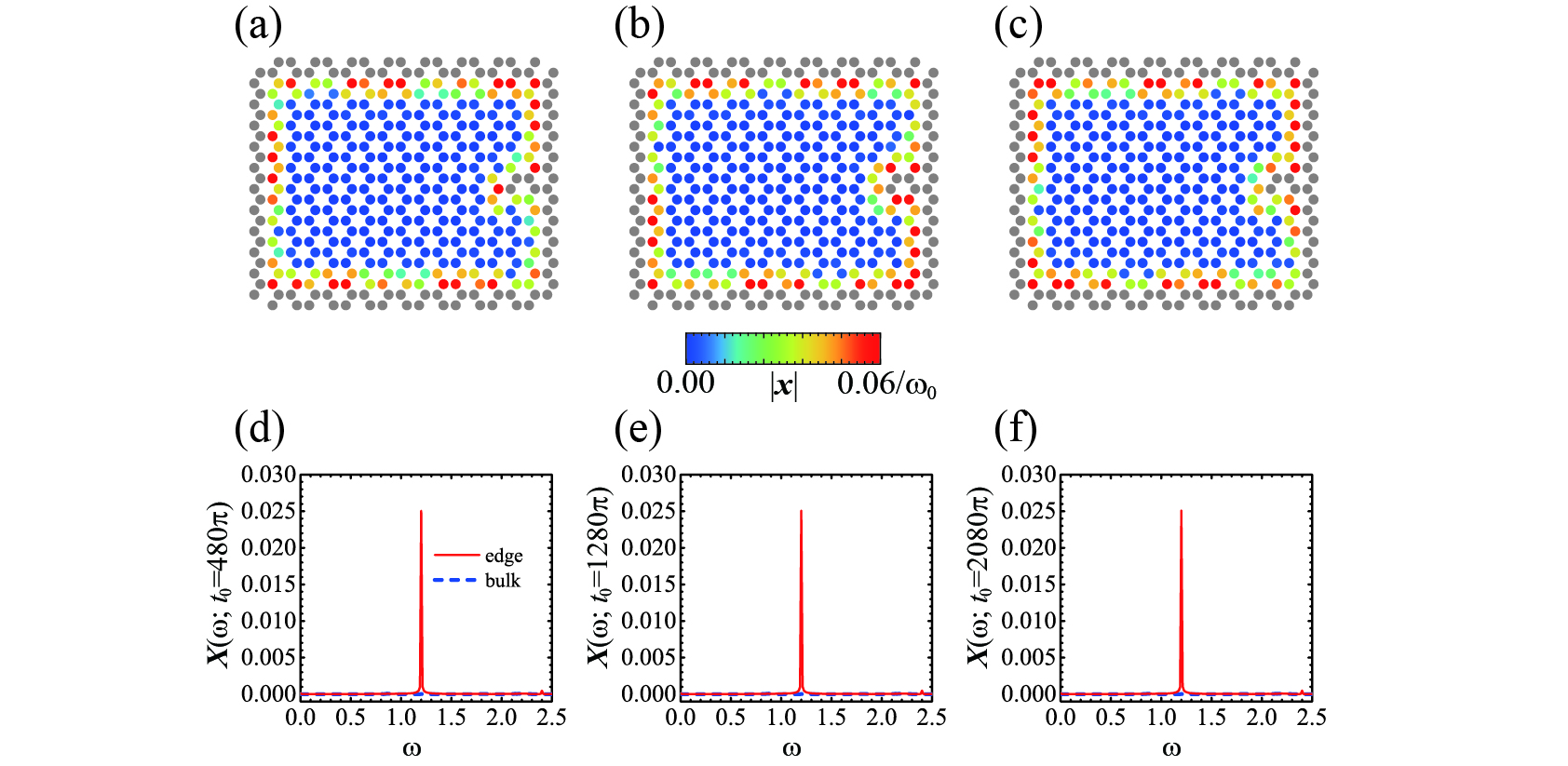}
    \caption{(a)-(c) Snapshots of evolution of the edge state with $\omega_0=1.20023$ at (a) $t=480\pi$, (b) $t=1280\pi$ and (c) $t=2080\pi$. (d)-(f) The edge and bulk vibration spectra of evolution of the edge state with $\omega_0=1.20023$ around (d) $t_0=480\pi$, (e) $t_0=1280\pi$ and (f) $t_0=2080\pi$.}
\end{figure}
\begin{figure}[ht]
    \centering
    \includegraphics[width=0.96\textwidth,angle=0]{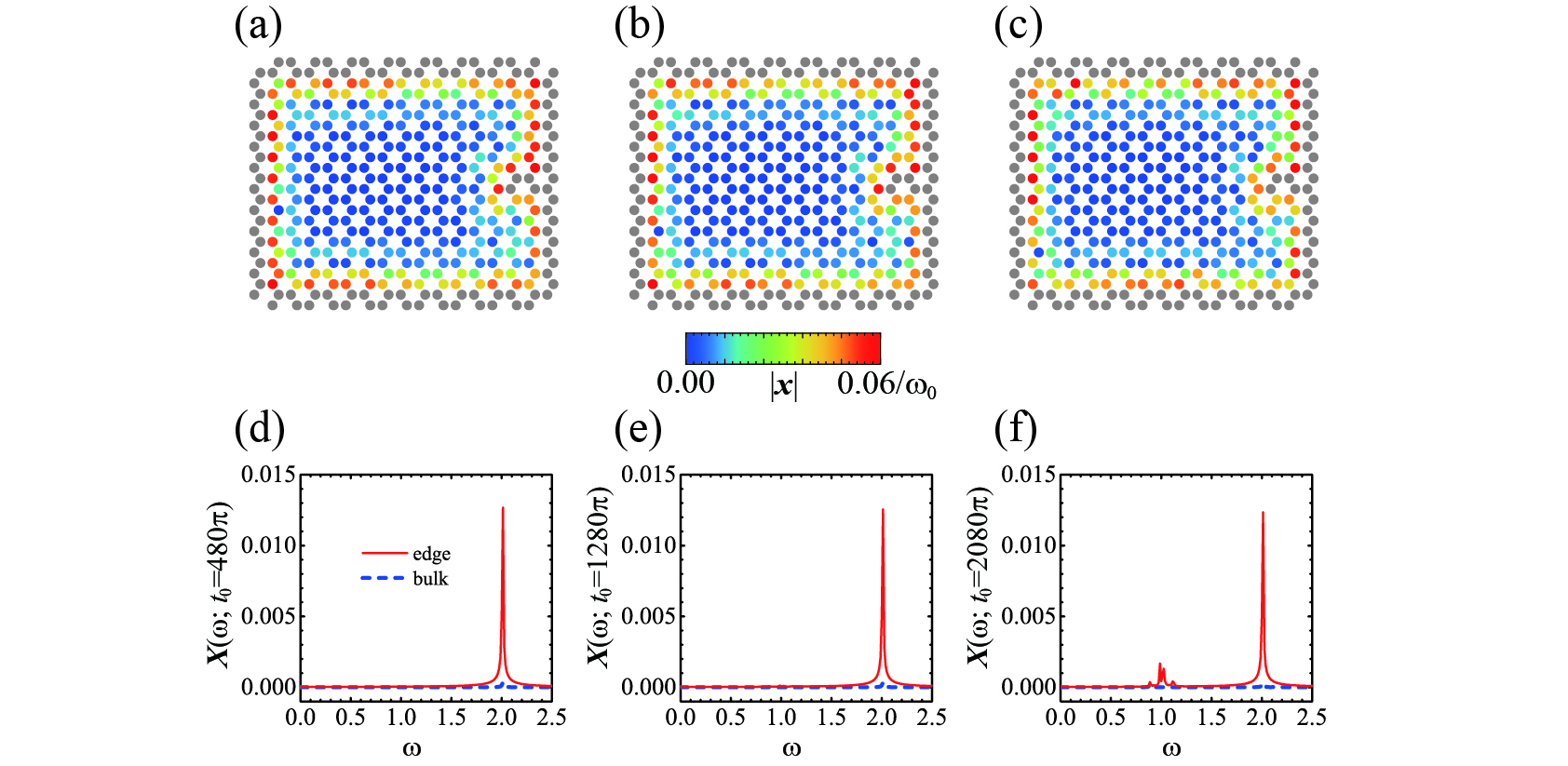}
    \caption{(a)-(c) Snapshots of evolution of the edge state with $\omega_0=2.00888$ at (a) $t=480\pi$, (b) $t=1280\pi$ and (c) $t=2080\pi$. (d)-(f) The edge and bulk vibration spectra of evolution of the edge state with $\omega_0=2.00888$ around (d) $t_0=480\pi$, (e) $t_0=1280\pi$ and (f) $t_0=2080\pi$.}
\end{figure}

\end{widetext}

\end{document}